\documentclass[12pt,preprint]{aastex}
\slugcomment{}

\shorttitle{Infrared Surface Brightnesses}

\shortauthors{Barnes et al.}

\begin{document}

\title{Infrared Surface Brightness Distances to Cepheids: a comparison of Bayesian and linear-bisector calculations}

\author{Thomas G. Barnes III}
\affil{The University of Texas at Austin, McDonald Observatory, 1 University Station,
C1402, Austin, TX 78712--0259;  tgb@astro.as.utexas.edu}

\author{Jesper Storm}
\affil{Astrophysikalisches Institut Potsdam, An der Sternwarte 16, D-14482 Potsdam, Germany; e-mail: jstorm@aip.de}

\author{William H. Jefferys}
\affil{The University of Texas at Austin, Dept. of Astronomy, 1 University Station,
C1400, Austin, TX 78712--0259;  bill@astro.as.utexas.edu}

\author{Wolfgang P. Gieren}
\affil{Universidad de Concepci\'on, Departamento de F\'isica, Casilla 160-C, Concepci\'on, Chile; e-mail: wgieren@coma.cfm.udec.cl}

\and

\author{Pascal Fouqu\'e}
\affil{Observatoire Midi-Pyr\'en\'ees, Laboratoire d'Astrophysique (UMR 5572),  14, avenue Edouard Belin, F-31400 Toulouse, France; e-mail: pfouque@ast.obs-mip.fr}

\begin{abstract}
We have compared the results of Bayesian statistical calculations and linear-bisector calculations  for obtaining Cepheid distances and radii by the infrared surface brightness method. We analyzed a set of 38 Cepheids using a Bayesian Markov Chain Monte Carlo method that had been recently studied with a  linear-bisector method. The distances obtained by the two techniques agree to $1.5\% \pm 0.6\%$ with the Bayesian distances being larger.  The radii agree to $1.1\% \pm 0.7\%$ with the Bayesian determinations again being larger.  We interpret this result as demonstrating that the two methods yield the same distances and radii.  This implies that the short distance to the LMC found in recent linear-bisector studies of Cepheids is not caused by deficiencies in the mathematical treatment.  However, the computed {\em uncertainties} in distance and radius for our dataset are larger in the Bayesian calculation by factors of 1.4--6.7.  We give reasons to favor the Bayesian computations of the uncertainties.  The larger uncertainties can have a significant impact upon interpretation of Cepheid distances and radii obtained from the infrared surface brightness method.

\noindent

\end{abstract}


\section{Introduction}

The infrared surface brightness technique is a powerful method for determining distances to Cepheid variables (Welch 1994\nocite{welch}, Fouqu\'e \& Gieren 1997\nocite{fg97}).  It is independent of other astrophysical distance scales, nearly independent of errors in reddening, and may be applied to arbitrarily-chosen, individual Cepheids at great distances.  However, the implementation of the infrared surface brightness technique and  its predecessor, the visual surface brightness method, have been criticized as not mathematically rigorous in their solutions to the surface brightness equations (Laney \& Stobie 1995, \nocite{ls} Barnes \& Jefferys 1999\nocite{bj}).  This leads to the possibility that the distances, radii, and their uncertainties may be erroneous. 

This is of current interest as Storm et al. (2004)\nocite{storm05} have found a short distance to the LMC based on a linear-bisector analysis of six Cepheids in the LMC cluster NGC1866.  They also found a slope to the LMC Cepheid period-luminosity relation that is substantially different than found in the OGLE magnitudes (Udalski et al. 1999)\nocite{udalski}.  It is important to determine if their results are affected by the mathematical method used in the surface brightness analysis. 

To address the larger issue Barnes et al. (2003) \nocite{mcmc} developed a Bayesian Markov Chain Monte Carlo (MCMC) solution to the surface brightness equations that {\em is} mathematically rigorous.  In the current paper we do a direct comparison for a significant sample of Cepheids between the Bayesian MCMC solution and recent  linear-bisector calculations (Storm et al. 2004\nocite{storm}, Gieren et al. 2005\nocite{gieren05}) to explore possible differences.  The two methods used identical data, identical surface brightness equations and identical physical constants to ensure that only the mathematical approaches were compared. 

In the next section we introduce the surface brightness method for determining distances and radii and present the background of the infrared surface brightness method.  We then discuss the data that are used for the two calculations.  We review the  linear-bisector calculations and the Bayesian MCMC calculations.  In section 6 we compare the results of the two methods for 38 Galactic Cepheids.  Finally we discuss the importance of the agreement and the differences that we find.  

\section{The Surface Brightness Method}

\subsection{The surface brightness equations}

The infrared surface  brightness method is a modification of the visual surface brightness technique developed by Barnes et al. (1977) \nocite{cepheids} and thus shares the same computational algorithm.   Because solution of the equations is the important issue, we introduce the equations in some detail.  Useful discussions of previous work have been given by Gieren, Barnes \& Moffett (1993), \nocite{gbm93} Fouqu\'e \& Gieren (1997), \nocite{fg97}  Nordgren et al. (2002), \nocite{nordgren2002} Fouqu\'e, Storm, \& Gieren (2003), \nocite{fsg03} and Barnes et al. (2003). \nocite{mcmc}

Barnes \& Evans (1976) \nocite{be} and Barnes, Evans, \& Parsons (1976) \nocite{bep} defined  a  visual surface brightness parameter $F_V$ as
\begin{equation}
F_V=4.2207-0.1V{\rm _0}-0.5\log\phi \label{eq:fparameter}
\end{equation}
\noindent
and also as,
\begin{equation}
F_V=\log{T_e}+0.1BC, \label{eq:equation 2}
\end{equation}
\noindent
where $V{\rm _0}$ is the stellar visual magnitude corrected for interstellar
extinction, $\phi$ is the stellar angular diameter expressed in milliarcseconds, $T_e$
is the effective temperature and $BC$ is the bolometric correction. 

They demonstrated that $F_V$ is well correlated with Johnson color index $(V-R){\rm _0}$ for a very wide range of stellar types.  \begin{equation}
F_V=A+B\left(V-R\right){\rm _0} \label{eq:basiceq}
\end{equation}
\noindent 
equation  (\ref{eq:basiceq}) is called the visual surface brightness relation. 

These relations led Barnes et al. (1977) \nocite{cepheids} to infer a distance scale for Cepheids as follows.  At each time $t$ in the pulsation of the Cepheid, equations (\ref{eq:fparameter}) and (\ref{eq:basiceq}) may be combined to obtain the angular diameter variation of the star, $\phi(t)$, 
\begin{equation}
4.2207-0.1V{\rm _0}(t)-0.5\log\phi(t)=A+B\left(V-R\right){\rm _0}(t)
\label{eq:combined}
\end{equation}
\noindent
In addition we infer the Cepheid's linear radius variation ${\Delta}R(t)$ about the mean radius from an integration of the radial velocity curve, $V_r(t)$, 
\begin{equation}
{\Delta}R(t)=-\int{p\left(V_r(t)-{V_\gamma}\right)dt}\label{eq:rvint}
\end{equation}
\noindent
where the factor $-p$ converts observed radial velocity to the star's pulsational velocity and $V_\gamma$ is the center-of-mass radial velocity of the star.  Integration of the discrete radial velocity data requires that the velocity variation be appropriately modeled. 

Substituting into equation (\ref{eq:combined}), the relation among mean angular diameter $\phi_0$, linear diameter $\Delta R(t)$, and distance $r$, we obtain
\begin{equation}
4.2207-0.1V{\rm _0}(t)-0.5\log(\phi_0+2000\Delta R(t)/r)=A+B\left(V-R\right){\rm _0}(t).
\label{eq:combined2}
\end{equation}
 \noindent
where $\phi_0$ is in milliarcseconds, $r$ is in parsecs, and $\Delta R(t)$ is in AU.  The factor $2000$
converts radius to diameter and arcseconds to milliarcseconds.

Until very recently direct solution to equation (\ref{eq:combined2}) for distance and diameter has proved daunting.  Various methods have been  tried, but for reasons given by Barnes \& Jefferys (1999),\nocite{bj} none of the methods is rigorous. 

Barnes et al. (1977)\nocite{cepheids}, Gieren et al. (1993, 1997)\nocite{gbm}\nocite{gfg97}  and Welch (1994)\nocite{welch} simplified the problem by solving equation  (\ref{eq:combined}) for $\phi (t)$, solving equation  (\ref{eq:rvint}) for $\Delta R(t)$ and then solving for $r$ and $\phi_0$ using ordinary least-squares solution on equation  (\ref{eq:simple})  taking $\Delta R(t)$ as the independent variable.
\begin{equation}
\Delta R(t)=r(\phi_0 + \phi (t))/2000\label{eq:simple}
\end{equation} 
\noindent
However, the least-squares calculations do not properly treat the {\em errors-in-variables} problem that arises from uncertainty in both $\Delta R(t)$ and $\phi (t)$.  The best that can be done using least-squares is the linear-bisector solution.  Isobe et al. (1990)\nocite{isobe}  showed that linear-bisector performs better than other least-squares solutions when the problem is symmetric in the variables, as is the case here. Storm et al. (2004)\nocite{storm} adopted the linear-bisector method.

Following Balona (1977),\nocite{balona} Laney \& Stobie (1995)\nocite{ls} linearized equation  (\ref{eq:combined2}) and applied a maximum-likelihood calculation to its solution.  They used an iterative maximum-likelihood method to solve for the larger amplitude Cepheids for which the linearization is invalid.  Their calculation addresses the  {\em errors-in-variables} problem. However, results from the maximum-likelihood method can be quite sensitive to accurate knowledge of the uncertainties in the data, as discussed by Gieren et al. 1997\nocite{gfg97}.  

All the above methods require that the observed radial velocities be modeled in order to do the integration of equation  (\ref{eq:rvint}).  This creates a {\em model selection} problem.  Most researchers choose a Fourier series to fit the radial velocities, but the number of terms to include in the series is subjectively chosen.  

Finally, whether least-squares or maximum-likelihood, the above  calculations fail to treat properly the propagation of observational error through the radial velocity integral, equation  (\ref{eq:rvint}).  Balona (1977)\nocite{balona} introduced a widely used approximation to the error in $\Delta R(t)$ based on the uncertainty in the radial velocity data and on the assumption that the radial velocity data are equally spaced in pulsation phase, which is rarely true.

It was to address these issues that the Bayesian MCMC method was developed by Barnes et al. (2003).\nocite{mcmc}  Their calculation correctly solves the {\em errors-in-variables} problem, objectively selects the number of terms in the Fourier series fits, and correctly propagates the observational error through the radial velocity integration.  Moreover, the Bayesian MCMC calculation does not demand that a particular Fourier series fit the data;  rather, each Fourier series that is fit to the data has a particular posterior probability.  That probability is used as a weight in determining the other quantities sought in the solution, i.e., distance, mean radius, etc.

A large number of calculations of variable star distances and radii by surface brightness methods are present in the literature.  It is therefore of interest to determine the extent to which the above listed deficiencies affect those results.  We will do this by comparing the linear-bisector calculations with Bayesian MCMC calculations. 

\subsection{The infrared surface brightness technique}

Welch (1994) \nocite{welch} first showed that use of the infrared combination $K{\rm _0}, (V-K){\rm _0}$  in place of $V{\rm _0}, (V-R){\rm _0}$ in equation  (\ref{eq:combined2}) has significant advantage in the precision of the {\em distances} and {\em radii} for Cepheids.  He found an improvement by a factor of three in the distance uncertainty for the Cepheid U Sgr.  Welch attributed the improved precision to several factors.  First, the color index $(V-K){\rm _0}$ is as good an indicator of surface brightness as bluer color indices but much less sensitive to the complications of line blanketing and surface gravity. Second, the $K{\rm _0}$ magnitude lies on the Rayleigh-Jeans tail of the flux distribution and is therefore less affected by the surface brightness variation.  It is obvious that a magnitude more sensitive to the radius variation that is then corrected for surface brightness variation using a color index that is a more accurate indicator of surface brightness ought to yield superior results.  

Laney \& Stobie (1995) \nocite{ls} independently examined two optical magnitude-color index combinations and two infrared combinations to determine which would give the most precise {\em radii} of Cepheids.  From both model atmosphere considerations and the precision of 49 Cepheid calculations, they concluded that $K{\rm _0}, (V-K){\rm _0}$ and $K{\rm _0}, (V-J){\rm _0}$ were superior to the optical indices.  (Examination of their Tables 5--6 suggests that $K{\rm _0}, (V-K){\rm _0}$ is slightly better, confirming Welch's choice.)   Laney \& Stobie also noted that the presence of a companion to the Cepheid would have less effect on infrared indices than on optical ones, as Cepheids are more likely to have a blue main-sequence companion than a red giant companion from stellar evolution considerations.  

Extending Welch's (1994) work on U Sgr, Fouqu\'e \& Gieren (1997) \nocite{fg97} compared distance and radius results  using $V{\rm _0}, (V-R){\rm _0}$; $V{\rm _0}, (V-K){\rm _0}$; and $K{\rm _0}, (J-K){\rm _0}$ with a somewhat improved data base and improved surface brightness equations.  They found the same distances and radii from all three combinations, but much superior precision for the infrared color indices.  The combination $V{\rm _0}, (V-K){\rm _0}$ seemed to be slightly preferred on the basis of precision.  When adjusted to the same surface brightness equations,  Fouqu\'{e} \& Gieren's results agree within the errors with those of Welch and of Laney \& Stobie.  

With a sample  of 16 Galactic cluster Cepheids, Gieren, Fouqu\'e, \& G\'omez (1997)  \nocite{gfg97} repeated the comparison of the above three combinations.  Again they concluded that the infrared color indices are superior to the optical in precision and also agree well with each other.  On the other hand, the $V{\rm _0}, (V-R){\rm _0}$ solutions gave distances and radii $\sim14\%$ larger than the infrared solutions.  While they indicated no preference between $V{\rm _0}, (V-K){\rm _0}$ and $K{\rm _0}, (J-K){\rm _0}$, their Table 3 shows that $V{\rm _0},(V-K){\rm _0}$ gave better precision than $K{\rm _0}, (J-K){\rm _0}$ in 13 of 16 cases.  

Based on these studies, we adopt the combination $V{\rm _0}, (V-K){\rm _0}$ as the basis for our comparison of the mathematical methods.  

\section{The Data}

Storm et al. (2004) \nocite{storm} analyzed $34$ Galactic Cepheids for distances and radii using the $V{\rm _0}, (V-K){\rm _0}$ infrared surface brightness technique. Gieren et al. (2005)\nocite{gieren05}  enlarged the sample by adding four more stars. We have used the full set of 38 Cepheids.  The infrared surface brightness relation adopted in both those studies and by us is the one determined by Fouqu\'e \& Gieren (1997): \nocite{fg97}

\begin{equation}
F_V=3.947 - 0.131 (V-K)_0\label{eq:ISB}
\end{equation}
\noindent

The individual stellar data required for the analyses are the photometric measures $V, (V-K)$, reddening $E(B-V)$, radial velocities $V_r$, pulsation period $P$, and pulsation phases $\theta$.  Storm et al. list the sources for photometry and radial velocities in their Table 1.  To ensure that we used identical data in both calculations, the Bayesian analysis used the same input data files as used in the linear-bisector analysis.  

We added Z~Lac, Y~Oph,  S~Sge, and CS~Vel  to the program using data referenced in Gieren et al. (2005)\nocite{gieren05}.  In addition, the results for $\ell$~Car given by Storm et al. were revised to incorporate new radial velocities, which are also referenced in Gieren et al. (2005)\nocite{gieren05}.  Again, the Bayesian analysis used the same input data files as used in the linear-bisector analysis by Gieren et al.

Although the infrared surface brightness method is largely independent of the interstellar extinction, it is important that the {\em same} extinction and reddening be used in our two calculations.  We follow the earlier studies and adopt $E(B-V)$ from Fernie's (1990) tabulation for Cepheids.  These $E(B-V)$ values are listed in Table 1 along with the periods of the Cepheids.  We adopted $A_V = 3.26 E(B-V)$ and $E(V-K) = 2.88 E(B-V)$, slightly different than the reddening law used by Storm et al.  The linear-bisector calculations were repeated with the new reddening law for all 38 Cepheids. 

In the integration of the radial velocity curve to obtain the linear displacements, equation  (\ref{eq:rvint}), the value of $p$ is required.  Some studies have adopted a single value of $p$ for all Cepheids, others have adopted individual values.  This choice has no effect upon our comparison of linear-bisector and Bayesian computations, provided the same value is used in both calculations for a specific Cepheid.  Storm et al. and Gieren et al. used a relation between $p$ and period developed by Gieren et al. (1989) \nocite{gbm} to approximate model atmosphere results by Hindsley \& Bell (1986): \nocite{hb}
\begin{equation}
p=1.39-0.03 log P
\end{equation}
\noindent 
where $P$ is the pulsation period. We adopt this relation as well,  using the periods given in Table 1 for the Bayesian MCMC calculation.  

\section{The Linear-bisector Computations}

Storm et al. (2004)\nocite{storm} and Gieren et al. (2005)\nocite{gieren05} applied a linear-bisector, least-squares solution to the surface brightness equations.   They solved equation  (\ref{eq:combined}) for the angular diameter variation $\phi (t)$.  The radial velocity data $V_r(t)$ were linearly interpolated in phase between the observed velocities and equation  (\ref{eq:rvint}) integrated in 0.01 steps in phase to obtain the $\Delta R(t)$ variation.   Finally, equation  (\ref{eq:simple}) was solved for $r$ and $\phi_0$. 

\begin{figure}
\plotone{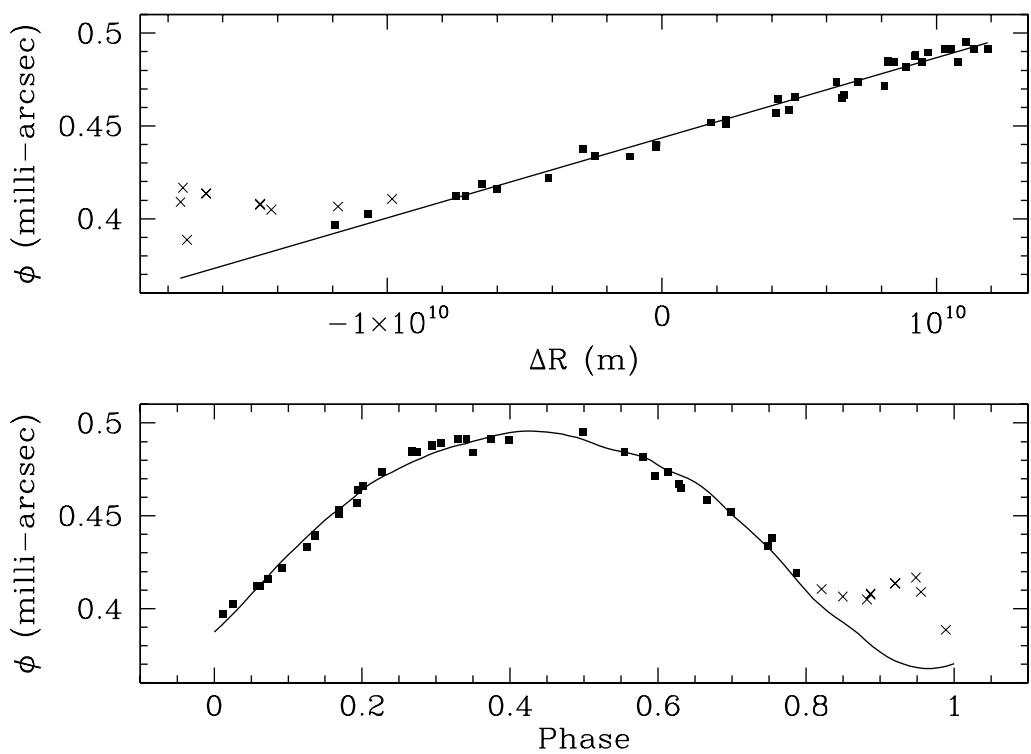}
\caption{Points represent the photometrically determined angular diameters for AQ Pup.  Crosses represent computed values that were eliminated from the fit. The line in the upper panel shows the bisector fit. The curve in the lower panel delineates the angular diameters obtained from integrating the radial velocity curve at the derived distance. } \label{stormfig2}
\end{figure}

To verify the solution for each Cepheid, the angular diameter variation and the displacement variation were plotted against phase in the pulsation cycle.   For many of the stars, the angular diameter variation was a very poor match to the displacement curve in the phase interval $0.8-1.0$, as illustrated in Figure \ref{stormfig2}.  Storm  et al. discussed the source of the poor fit without coming to a conclusion and decided that the best course of action was to ignore this phase interval in the fit.  They deleted the phase interval $0.8-1.0$ for all stars in their fit of equation  (\ref{eq:simple}) for $r$ and $\phi_0$. 

Some Cepheids showed a small phase shift between the photometry and the radial velocities.  The phase shift causes a loop in the upper panel and a displacement between the fitted curve and the points in the lower panel.  For the stars showing this effect a phase shift was determined by minimizing the scatter in the upper panel. The phase shift has already been imposed between the photometry and radial velocities used for Fig. \ref{stormfig2} and thus these effects are not seen.  The adopted phase shifts are listed in Table 1.  

There are two constants adopted in a surface brightness calculation that must be the same in the linear-bisector and Bayesian calculations.  We chose to fix the constants in the Bayesian calculation to those used in the Storm et al. paper.  These constants are the constant term in the surface brightness definition, equation  (\ref{eq:fparameter}), and the conversion factor from angular diameter (in milliarcseconds) and distance (parsecs) to linear radius (solar radii).  The first of these was taken to be 4.2207 and the second as  0.10727 solar radii per mas-parsec (see eq. (\ref{eq:simple})). 

To perform the calculations Storm et al. \nocite{storm} used the FORTRAN subroutine SIXLIN which is  available from Isobe et al. (1990)\nocite{isobe}. The computations were run on a Linux personal computer and took a fraction of a second per star.

The quantities determined in the linear-bisector calculation that are of interest to us here are the distance $r$, the mean linear radius $R$, computed from the distance and mean angular diameter $\phi_0$, and their 1 $\sigma$ uncertainties.  These quantities are given in Table 1.  
 
\section{The Bayesian Markov Chain Monte Carlo Computations}

The Bayesian MCMC method that we applied here  is described in detail by Barnes et al. (2003)\nocite{mcmc}.  It is important to realize that the Bayesian calculation treats the {\em unknowns} in the problem as {\em probability distributions}, not as specific values to be determined.  The goal of the analysis is to determine the probability distribution for each parameter of interest, from which inferences may be drawn by appropriate means and variances.

To compute the  {\em posterior probability distribution} for each parameter requires the likelihood function on a specific model, appropriate priors, and sampling strategies for all parameters.  The {\em likelihood function} is the probability of obtaining the particular data given the model. Bayesian statistics encapsulate our understanding of the parameters in the model, prior to considering the data, in {\em prior distributions}.  The {\em posterior probability distribution} is the product of the prior and the likelihood, appropriately normalized.  

The full posterior probability distribution in this problem requires solution to integrals in the normalization that cannot be done analytically.   However, the unnormalized posterior probability distribution avoids these integrals and can be used to generate a Monte Carlo sample from the full posterior probability distribution.  The techniques used to generate the sample are Markov Chain techniques. 

The model for the infrared surface brightness calculation is developed by first substituting the infrared color index for the visual color index in equation  (\ref{eq:combined2}) and then rearrange as follows
\begin{equation}
\left(V-K\right){\rm _0}(t)=\frac{1}{B}\left(4.2207-0.1V{\rm _0}(t)-A-0.5\log\left(\phi_0+2000\Delta
R(t)\pi\right)\right) \label{eq:likelihood}
\end{equation}
\noindent
where $A,B$ take the values given in equation  (\ref{eq:ISB}) and where we have replaced $1/r$ with $\pi$, the parallax in arcseconds.   Within this model the likelihood function is specified in a straightforward way and is given in equation (12) of Barnes et al. (2003)\nocite{mcmc}.  We model the photometry and the radial velocity data as drawn from normal distributions with variances given by the observational uncertainties.  Because we do not trust the quoted observational uncertainties, we introduce a hyper-parameter scale factor on each variance to model deviation in the scatter from that expected from the quoted uncertainties.  The time variations of the photometry and radial velocities are modeled by Fourier series of unknown order on the pulsation phase. 

Barnes et al. discuss the priors adopted for each parameter of interest, but only one is relevant here.  It is well-known that Cepheids are distributed within the plane of the Galaxy with an exponential decrease in density away from the plane.  We adopted a prior on distance that reflects the flattened density distribution with a scale height of $70 \pm 10$ pc.  The results are insensitive to reasonable changes in the scale height. 

The sampling strategy employed in this work is Markov Chain Monte Carlo using the Metropolis-Hastings and Gibbs algorithms, as described in Barnes et al.  The art in this approach is to find sampling methods that explore the posterior probability distributions fully and efficiently.  Internal tests provide guidance on the completeness and efficiency of the sampling.  All the results presented here passed those tests.   In the customary manner, we chose a {\em burn-in} phase to improve the model selection efficiency.   

To be consistent with the linear-bisector calculation, we deleted from the Bayesian solution the photometry and radial velocities in pulsation phase interval $0.8-1.0$.  In the general model, we allow for an unknown phase shift between the photometry and the radial velocities; in this calculation, we fixed the phase shift at the value determined by the linear-bisector calculation.   

The calculations were run on a 1 GHz Macintosh G4 computer under system MacOS 10.3 using the statistical language R-1.8.0$\beta$ distributed by the R Development Core Team.\footnote{See http://www.r-project.org/ .} (The R language is available for other platforms.)  We used a burn-in of 1,000 samples followed by 10,000 samples.  Because the R code is interpreted code, the calculations run slowly; the computations here typically took about an hour per star. 

Posterior probability distributions were determined for all the model parameters of interest. These are the parallax $\pi$, the mean angular diameter $\phi_0$, the orders of the Fourier series on the apparent magnitudes and the radial velocities, the aforementioned hyper-parameters on the observational uncertainties, the mean $V{\rm _0}$ magnitude (both intensity mean and magnitude mean), and the center-of-mass radial velocity.  For our purpose here, only the parallax and mean angular diameter are important. These were converted to distance and mean linear radius within the code and listed with their uncertainties  in Table 1. (The erratum published by Barnes et al. (2003)\nocite{mcmc} was addressed in the computation of the radii.) 

\section{Comparison of the Results}

\subsection{The distances and radii}

\begin{figure}
\plotone{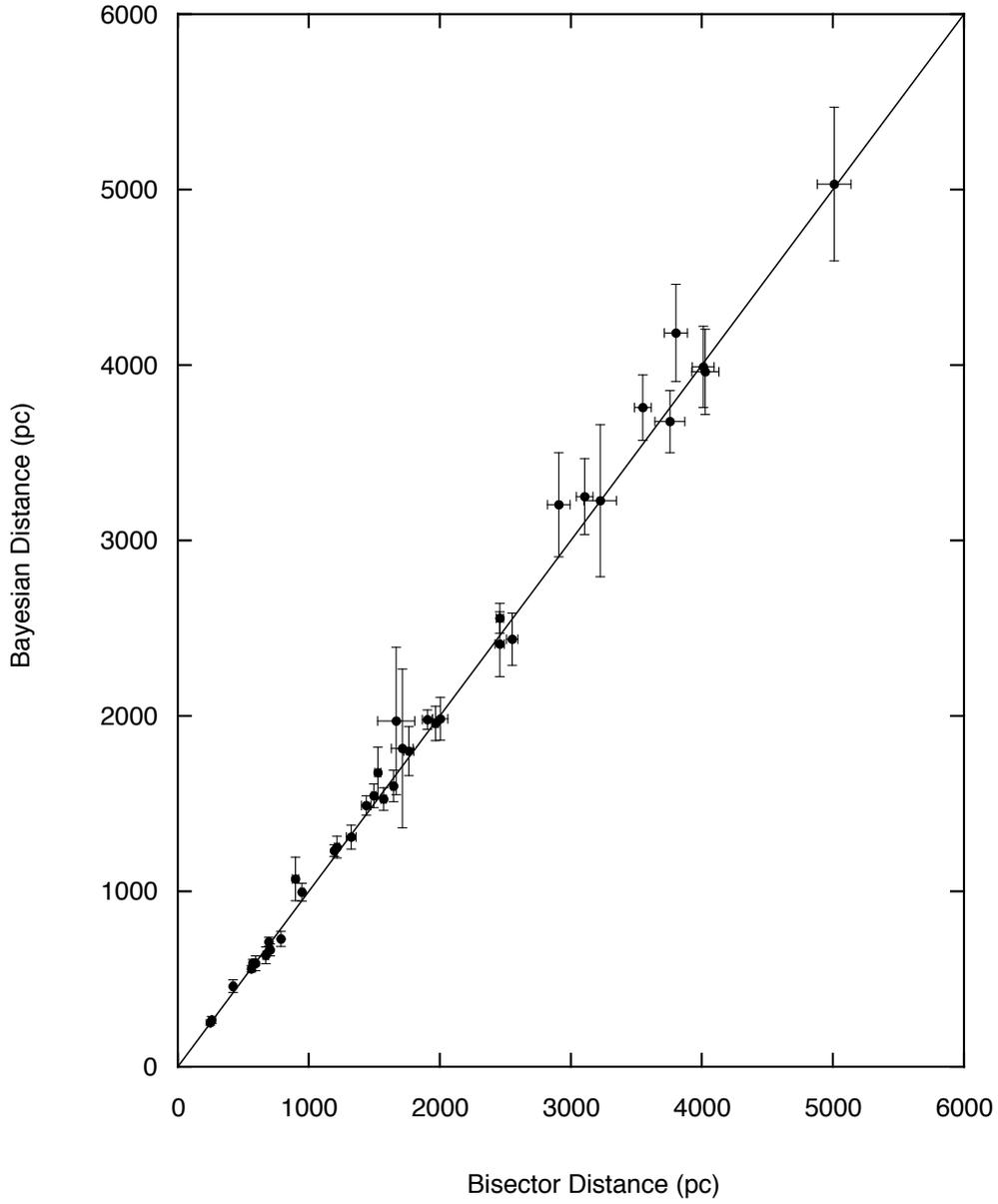}
\caption{The Bayesian distances are plotted against the linear-bisector distances, each with its 1$\sigma$ error bar. For nearby Cepheids the uncertainties are smaller than the symbols. The identity line is shown for reference. } \label{distances}
\end{figure}

\begin{figure}
\plotone{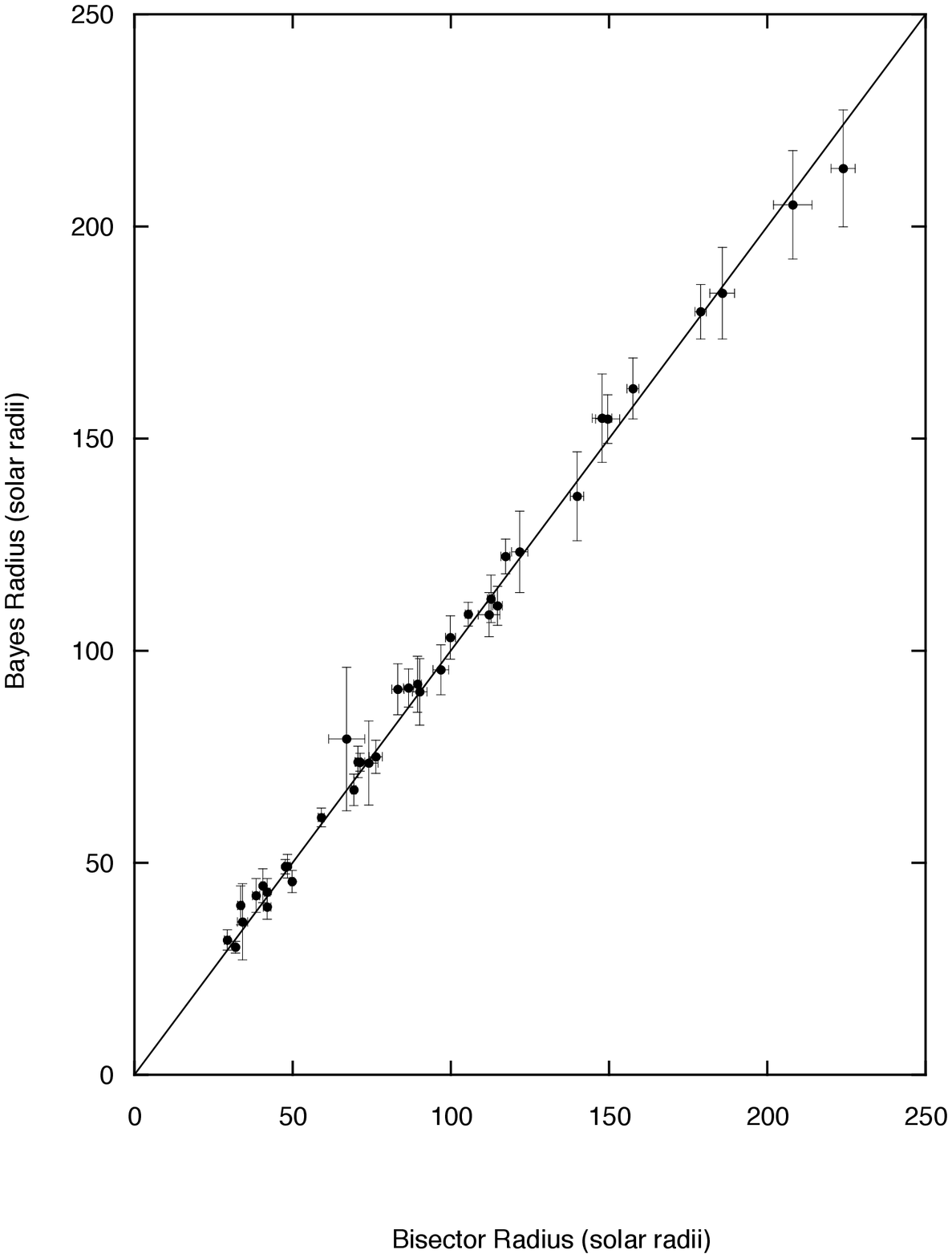}
\caption{The Bayesian radii are plotted against the linear-bisector radii, each with its 1$\sigma$ error bar. For smaller Cepheids the uncertainties are comparable to the symbols in size. The identity line is shown for reference. } \label{radii}
\end{figure}

\begin{figure}
\plotone{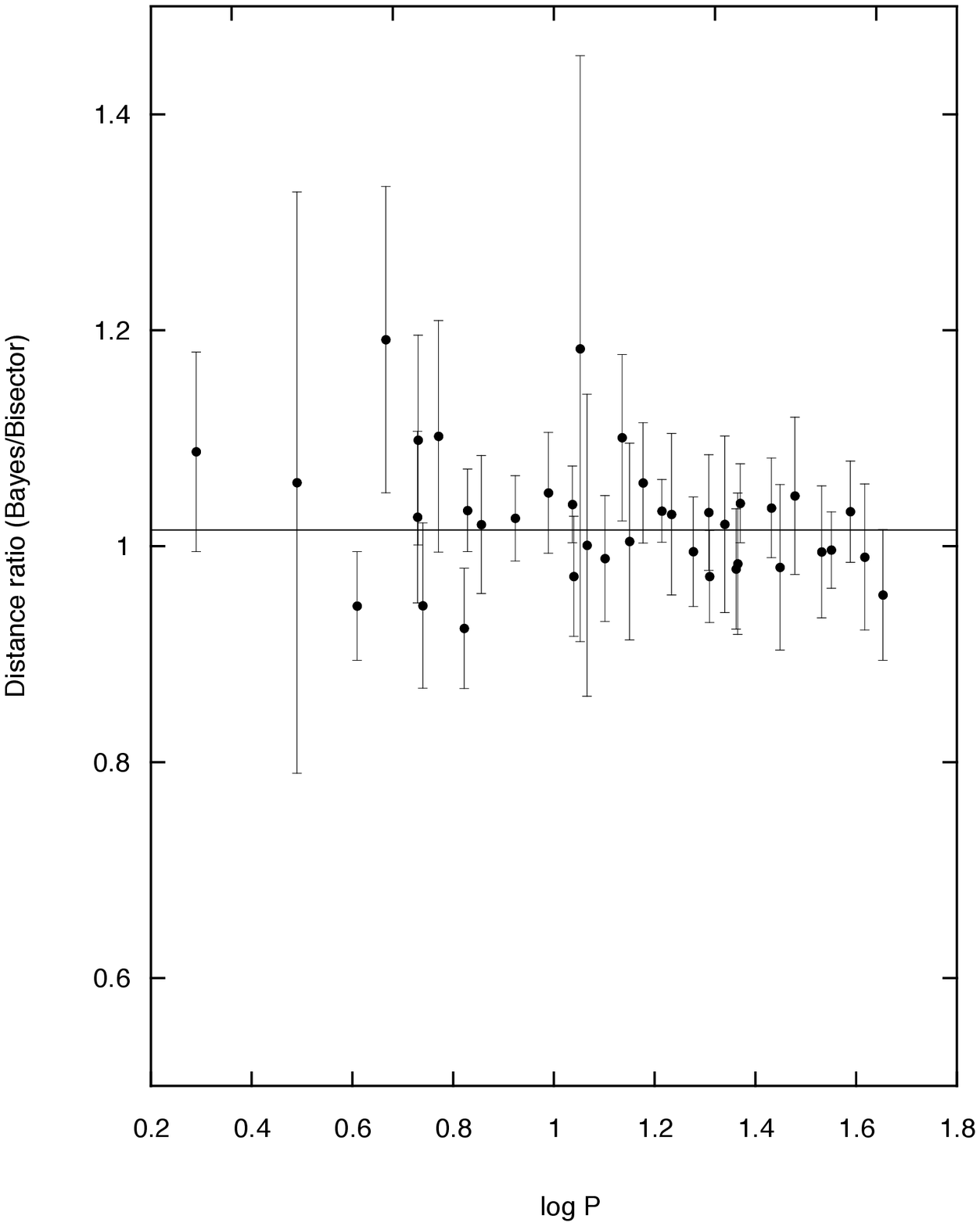}
\caption{The ratio of Bayesian distance to bisector distance is plotted against $log P$. The combined uncertainty is shown as a 1$\sigma$ error bar. The line at $ratio = 1.016 \pm 0.006$ is the weighted mean value. } \label{distratio}
\end{figure}

Our goal is to determine differences in the results of the two calculations with respect to distance and radius.   Figures \ref{distances} and \ref{radii} show the Bayesian values plotted against the linear-bisector values.  There is no obvious difference between the values from the two calculations. To look for subtle differences, we computed the ratio of the Bayesian distance to the linear-bisector distance.  A weighted least-squares fit of this ratio against $log (P)$ gives (Figure \ref{distratio})

\begin{figure}
\plotone{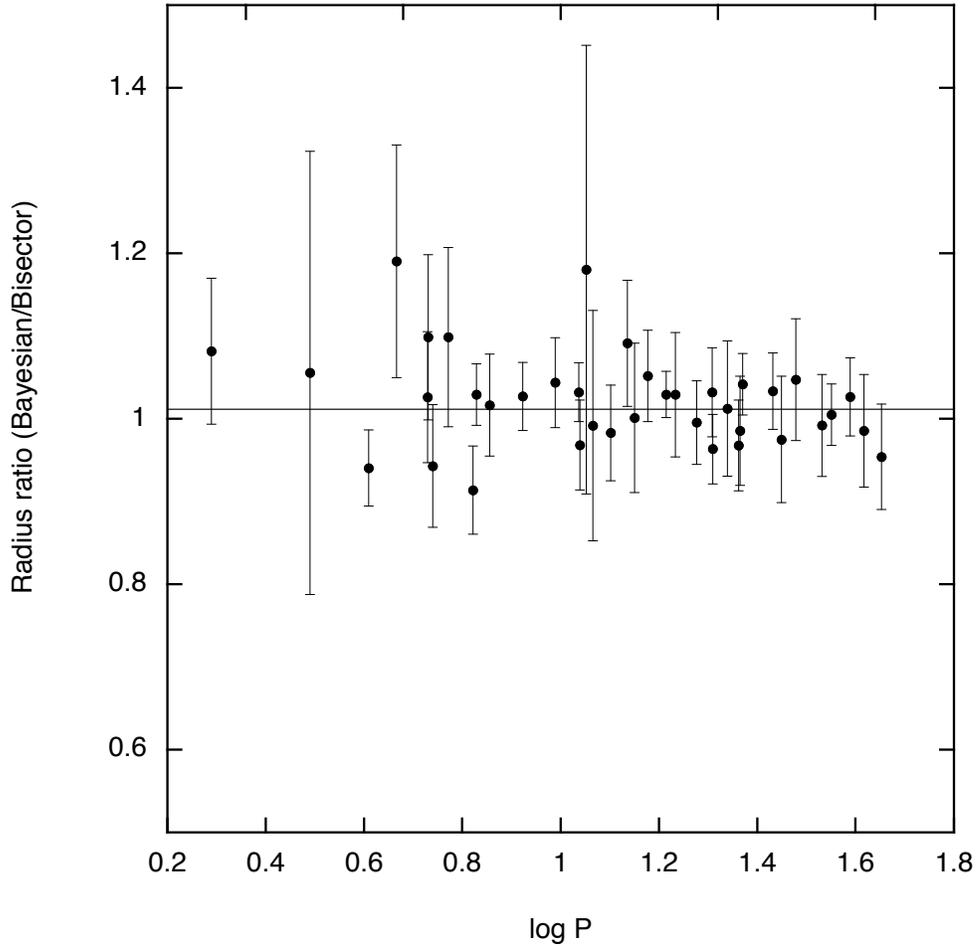}
\caption{The ratio of Bayesian radius to bisector radius is plotted against $log P$. The combined uncertainty is shown as a 1$\sigma$ error bar. The line at $ratio = 1.012 \pm 0.007$ is the weighted mean value. } \label{radratio}
\end{figure}

\begin{equation}
Ratio(r) = 1.016(\pm0.007)-0.012(\pm0.022)(logP - 1.113)
 \end{equation}
 
 \noindent
 where the fit is centered on the mean period. A weighted fit for the radius ratio gives (Figure \ref{radratio})
 
 \begin{equation}
Ratio(R) = 1.012(\pm0.007)-0.004(\pm0.023)(logP - 1.113)
 \end{equation}
 
\noindent
These show no evidence for any dependence on pulsation period.  Therefore we take a weighted mean for each ratio: the  distance calculations differ by $1.5\%\pm0.6\%$, with the Bayesian being larger, and  the radius calculations differ by $1.1\%\pm0.7\%$, with the Bayesian again being larger.  For comparison, the best individual distance and radius measurements in our dataset are for X Cyg, for which the Bayesian uncertainties are both $\pm2.6\%$, larger than the possible systematic difference between the Bayesian and bisector calculations. Our first result is that the distances and radii computed by the two methods agree quite well. 

\subsection{Uncertainties in distance and radius}

In Figures  \ref{distratio} and \ref{radratio} the uncertainties can be seen to increase as the period is shorter.  This is likely a result of the smaller pulsation amplitudes at shorter periods, which result in the photometric and velocity uncertainties having greater effect  on the computed distances and radii. 

More importantly, a  glance at Table 1, or Figures \ref{distratio} and \ref{radratio}, shows that the uncertainties in the distances and radii disagree substantially between the two calculations.  Typically the Bayesian uncertainty is more than three times the linear-bisector uncertainty.  

\begin{figure}
\plotone{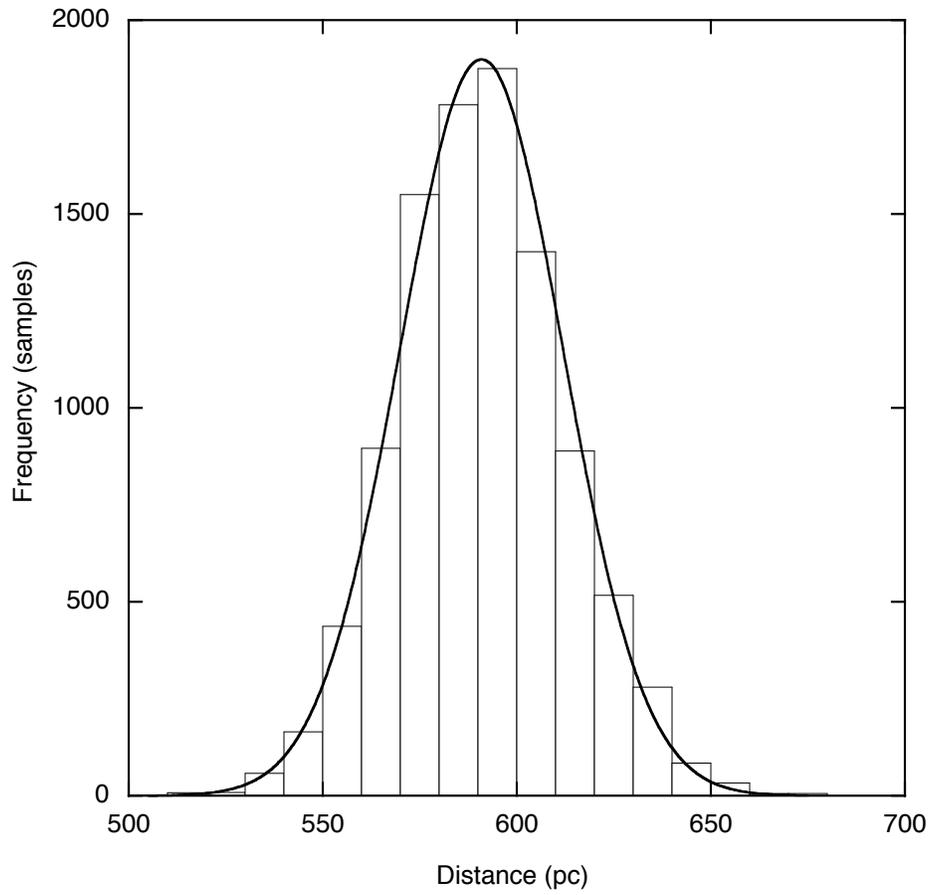}
\caption{The posterior probability distribution for the distance to U Sgr is shown as a histogram.  A normal distribution of the same mean distance, sigma and area is over-plotted. } \label{histogram}
\end{figure}

Because the concept of  an uncertainty estimated from the Bayesian MCMC posterior probability distribution  may not be clear, we show an example in Figure \ref{histogram}, the posterior probability distribution for the distance to U Sgr, which star is typical of our results.  Over-plotted on the probability distribution is a normal distribution constructed for the same distance (592 pc), sigma ($\pm21$ pc) and area (10,000 samples). The normal distribution describes the posterior probability distribution for the distance very well.  This justifies our adopting the sigma of the corresponding normal distribution as a 1$\sigma$  estimator for the uncertainty in the distance (and similarly for the radius) determined in the Bayesian calculation.  This estimator is then compared to the 1$\sigma$ estimator from the linear-bisector computation.

We add a caveat to the previous paragraph.  Because our computation determines the stellar parallax, not the stellar distance, (see equation (\ref{eq:likelihood})) the posterior probability distribution for the distance can become  asymmetric when the errors are large.  As the uncertainties become large, the parallax posterior probability distribution becomes broad (large sigma).  Its reciprocal, the distance posterior probability distribution, will also become broad {\em and necessarily asymmetric to larger distances}.  The same asymmetry will arise for the radius posterior probability distribution in such cases because that distribution is the product of the angular diameter posterior probability distribution (symmetric) and the distance posterior probability distribution (asymmetric). Note that these asymmetric distributions are {\em real} and not mathematical artifacts; they properly represent our knowledge of the distance and radius, which is not true for least-squares or maximum-likelihood calculations on the same data.  The latter methods assume symmetric errors by their very nature.  Because this situation prevails for only a few stars in this sample, and only for stars with large errors, it has little effect on the weighted mean ratios of distances and radii quoted in the previous section.  

\begin{figure}

\plotone{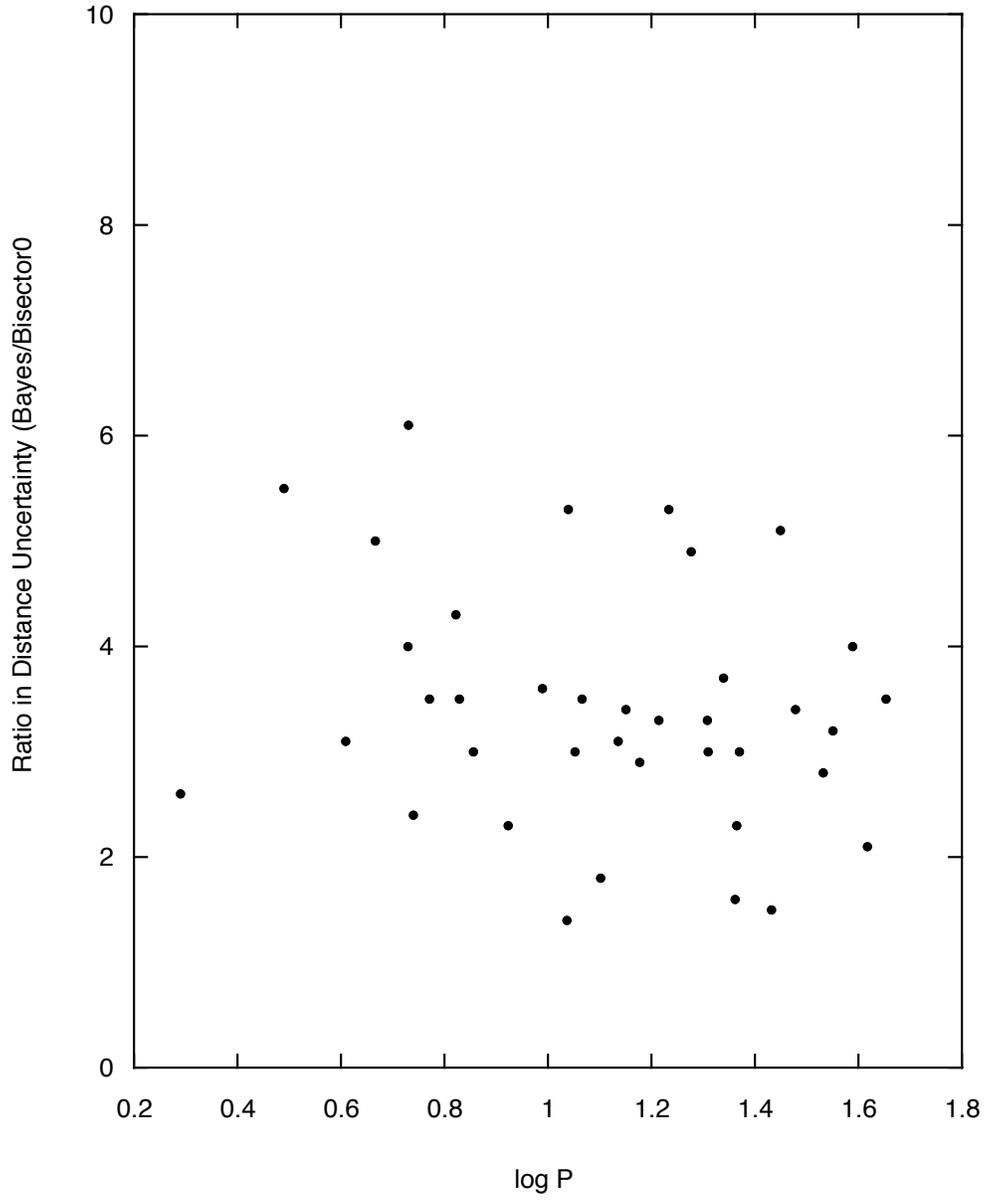}
\caption{The ratio of the Bayesian distance uncertainty to the bisector distance uncertainty for the same Cepheid is plotted against $log P$.} \label{sigmadist}
\end{figure}

\begin{figure}
\plotone{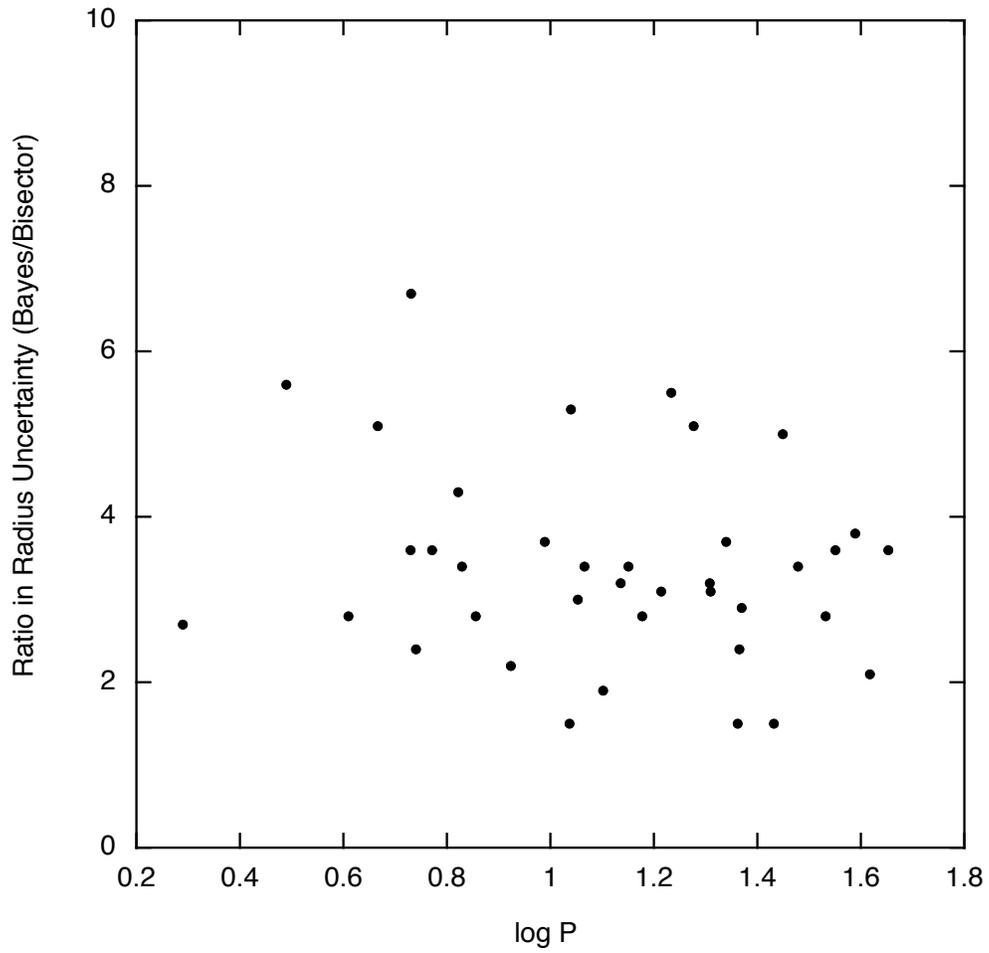}
\caption{The ratio of the Bayesian radius uncertainty to the bisector radius uncertainty for the same Cepheid is plotted against $log P$. } \label{sigmarad}
\end{figure}

As we did with the distances and radii themselves, we begin by examining the behavior of the {\em ratio} of the Bayesian uncertainty to the linear-bisector uncertainty  for the same Cepheid.  In Figures \ref{sigmadist} and \ref{sigmarad} we show these ratios for the distance and radius uncertainties plotted against $log P$.   Unweighted least-squares fits in these figures yield
 \begin{equation}
Ratio(r) = 3.396(\pm0.183)-0.748(\pm0.549)(logP - 1.113)
 \end{equation}
 \begin{equation}
Ratio(R) = 3.281(\pm0.202)-0.673(\pm0.606)(logP - 1.113)
 \end{equation}
\noindent
There is no apparent dependence of these ratios on pulsation period. Plots of the ratios of the distance uncertainties against distance and of the ratios of the radius uncertainties against radius are similarly uninformative. 

\begin{figure}
\plotone{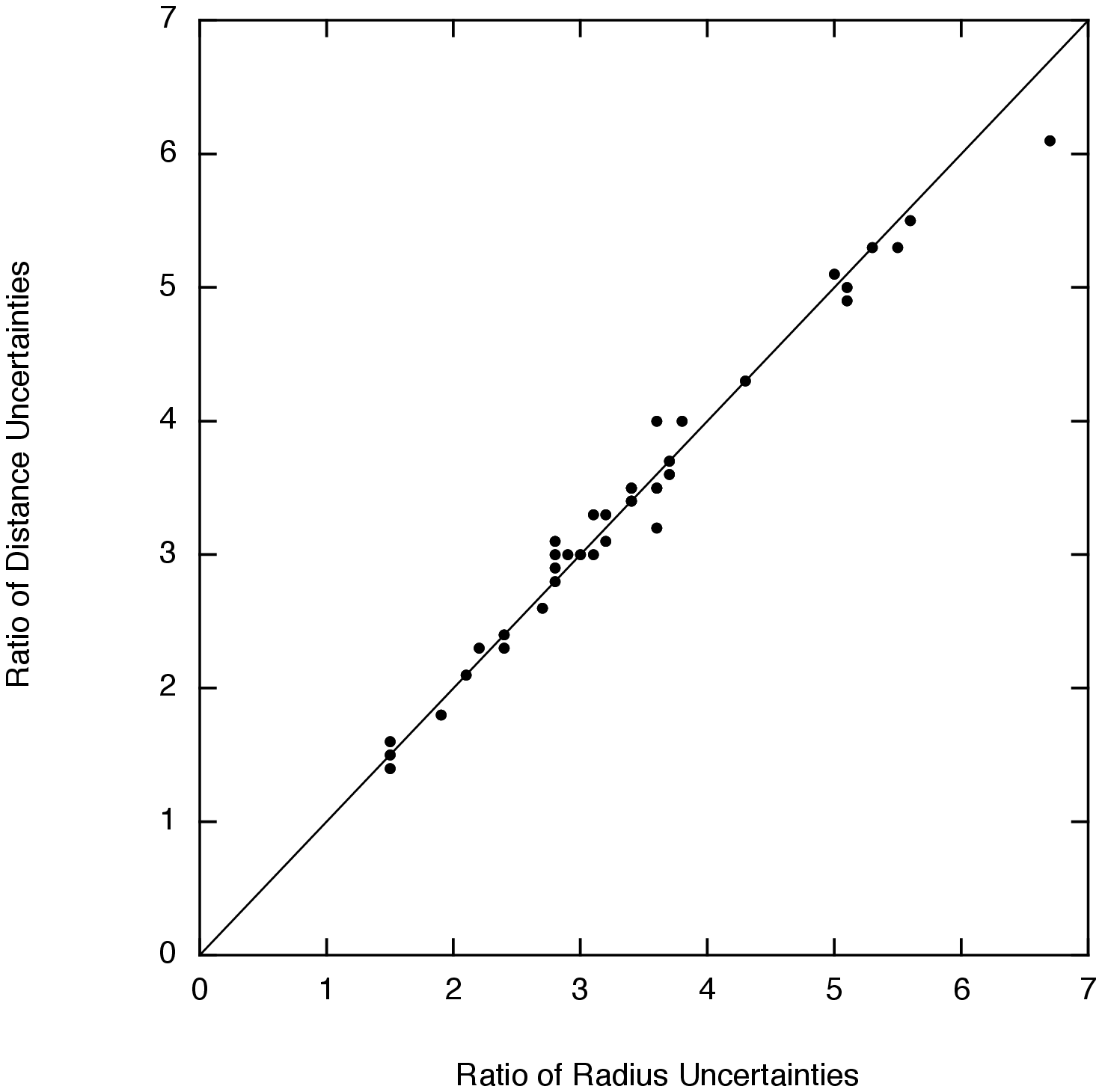}
\caption{The ratio of the Bayesian distance uncertainty to the bisector distance uncertainty is plotted against the ratio of the Bayesian radius uncertainty to the bisector radius uncertainty. The identity line is shown for reference. } \label{sigma_sigma}
\end{figure}

The two ratios are, however, highly correlated with each other ($R = 0.99$) as shown in Figure \ref{sigma_sigma}.  Thus the underlying cause of the larger uncertainties in the Bayesian calculation is likely to be the same for the distance uncertainty and radius uncertainty.  

In section 2.1 we noted that the linear-bisector calculation does not  treat the {\em errors-in-variables} problem rigorously nor does it properly propagate uncertainty through the radial velocity integration.  The second of these issues will certainly lead to an underestimate of the uncertainties in the computed distances and radii.  Because the Bayesian MCMC calculation {\em does} correctly address these two computational issues, we interpret the large ratio of Bayesian to bisector uncertainty as measuring the amount by which the linear-bisector errors have been underestimated.  This interpretation is supported by the fact  that {\em none} of the linear-bisector uncertainties is {\em larger} than its Bayesian counterpart. Our second result is that the linear-bisector calculation underestimates the uncertainties in distance and in radius substantially, amounting to factors of 1.4--6.7 for this dataset. This large range implies that the ratio that is obtained depends on the specifics of the data for the Cepheid which varies from star to star.  

\section{Discussion}

We set out to determine whether infrared surface brightness estimates of Cepheid distances and radii by the linear-bisector calculation are affected by the known mathematical shortcomings of that calculation.  Based on comparison of Bayesian MCMC and linear-bisector calculations for 38 Cepheids using the same data, same surface brightness equations, and same physical constants, we find that the distances and radii are {\em not} adversely affected but that the uncertainties in these quantities are seriously underestimated in the linear-bisector calculation. 

We find that Cepheid distances determined by the two calculations agree to $1.5\%\pm0.6\%$ with the Bayesian distances being larger.  This may be compared to the smallest individual uncertainty in distance found in the Bayesian calculation of $\pm2.6\%$.  Similarly we find that Cepheid radii determined by the two calculations agree to $1.1\%\pm0.7\%$ with the Bayesian radii being larger.  This may be compared to the smallest individual uncertainty in radius found in the Bayesian calculation of $\pm2.6\%$.  Any systematic difference between the mathematical approaches is both smaller than a $2\sigma$ effect and smaller than the typical single-star (Bayesian) uncertainty dictated by the data. 

These results have an impact on interpretation of infrared surface brightness results for Cepheids. For example, Storm et al. (2005)\nocite{storm05} used six Cepheids in an LMC cluster to infer a distance to the LMC of $(m-M)_0 = 18.30\pm0.07$ mag. by means of a linear-bisector solution for the infrared surface brightness equations.  This is less than the generally accepted distance of 18.50 mag.  From the present work, we can say that the linear-bisector calculation used by Storm et al. is {\em not} the cause of the smaller distance modulus. 

Similarly, that work found a much smaller slope for the Cepheid PL relation in the LMC than had been found by the OGLE project.  From the absence of a period dependence between the Bayesian MCMC distances and the linear bisector distances (Fig. \ref{distratio}) and from the overall agreement between the two,  we can be certain that the smaller slope found by Storm et al. (2005)\nocite{storm05} is {\em not }a result of using the linear-bisector, least-squares method.  

The uncertainties in distance and radius determined by the Bayesian MCMC calculation are much larger than determined by the linear-bisector calculation.  Given the known problems in a least-squares solution to the surface brightness equations, we interpret this as measuring the amount by which the linear-bisector computation underestimates the uncertainties.   It is important to note that the ratio of Bayesian to linear-bisector uncertainty ranges from 1.4 to 6.7 in this set of 38 Cepheids.  Clearly the amount by which the linear-bisector method underestimates the uncertainty depends on the specific nature of the data.  This is expected, but inconvenient, as it is not possible to simply multiply published linear-bisector uncertainties by a constant correction factor.  

Our result on the underestimation of the uncertainties in a linear-bisector calculation is supported in the previously mentioned paper by Storm et al. (2005)\nocite{storm05}. As the six Cepheids studied by them are in an LMC cluster, we can be confident that they are at the same distance.  The scatter in their distances is then an estimate of the true uncertainty in the linear-bisector distance to the cluster.  This scatter was found by Storm et al. (2005)\nocite{storm05} to be twice the formal errors of the linear-bisector distances to the Cepheids, within the range of results determined here. 

Other mathematical approaches have been used to solve the surface brightness equations for distance and radius.  Ordinary least-squares assumes no error on one variable and all errors on the other. Inverse fit least-squares assumes the reverse. The limitations of ordinary linear least-squares solutions (direct and inverse) include not only underestimation of the errors, but also possible systematic bias in the resulting distances and radii as discussed by Laney \& Stobie (1995)\nocite{ls} and Gieren et al. (1997)\nocite{gfg97}.  

The linear-bisector, least-squares calculation achieves a solution in-between the two other least-squares calculations  (direct and inverse).  Moreover, the linear-bisector error bar roughly corresponds to the difference between the two least-squares solutions.  Maximum likelihood uses information on errors on one or both variables to choose a result between the two results of linear least-squares (direct and inverse fits).  As a result, maximum likelihood cannot differ by more than one linear-bisector sigma from ordinary least-squares and even less from a linear-bisector fit.  As we have shown that this linear-bisector sigma is about 1/3 of the Bayesian sigma, the maximum likelihood results should also be about 1/3 of the Bayesian values. Our results for the linear-bisector solutions thus suggest that the maximum likelihood method would yield distances and radii in the infrared surface brightness method that are unbiased if the uncertainties in the data are well understood.  Unfortunately, these uncertainties are often not well understood. First, Barnes et al. (2003)\nocite{mcmc} showed that quoted uncertainties in Cepheid photometry and radial velocities are usually underestimated. Second, maximum-likelihood calculations usually adopt an approximation for the uncertainty in the displacements advocated by Balona (1977)\nocite{balona} for equally spaced velocity data. Our work demonstrates that this approximation does not apply to typical unequally-spaced radial velocity curves; if it did apply, the linear-bisector method would have yielded uncertainties in distance and radius close to those of the Bayesian MCMC calculation.  Thus we expect the maximum likelihood {\em uncertainties} to be underestimated as are the linear-bisector uncertainties and for the same reasons.

\acknowledgments
TGB and WHJ gratefully acknowledge financial support for this work  from McDonald Observatory and the Department of Astronomy of the University of Texas at Austin.  WPG acknowledges financial support for this work from the Chilean Center for Astrophysics FONDAP 15010003.

\newpage

\begin{deluxetable}{rclr|rrrr|rrrr}
\rotate
\tablenum{1}
\tablewidth{7.5in}
\tablecaption{Distance and Radius Results}
\tablehead{\colhead{Cepheid} &\colhead{Period} &\colhead{E(B-V)} &\colhead{$\Delta \theta$} &\colhead{r} &\colhead{$\sigma_r$} &\colhead{r} &\colhead{$\sigma_r$} & \colhead{R} & \colhead{$\sigma_R$} & \colhead{R} & \colhead{$\sigma_R$}\\ & days & mag & phase & pc & pc & pc & pc & $R_{\odot}$ & $R_{\odot}$ & $R_{\odot}$ & $R_{\odot}$\\ \hline & & & & LSQ & & Bayes & & LSQ & & Bayes}
\startdata
SU Cas & 1.949322 & 0.287 & 0.000 & 423 & $\pm$14 & 460 & $\pm$36 & 29.4 & $\pm$0.9 & 31.8 & $\pm$2.4\\
EV Sct & 3.090990 & 0.679 & 0.045 & 1714 & $\pm$83 & 1815 & $\pm$453 & 34.2 & $\pm$1.6 & 36.1 & $\pm$9.0\\
BF Oph & 4.067510 & 0.247 & 0.035 & 705 & $\pm$11 & 666 & $\pm$34 & 32.0 & $\pm$0.5 & 30.1 & $\pm$1.4\\
T Vel & 4.639819 & 0.281 & 0.000 & 899 & $\pm$25 & 1071 & $\pm$124 & 33.6 & $\pm$0.9 & 40.0 & $\pm$4.6\\
$\delta$ Cep & 5.366341 & 0.092 & 0.000 & 260 & $\pm$5 & 267 & $\pm$20 & 42.0 & $\pm$0.9 & 43.1 & $\pm$3.2\\
CV Mon & 5.378793 & 0.714 & 0.015 & 1527 & $\pm$24 & 1677 & $\pm$146 & 40.6 & $\pm$0.6 & 44.6 & $\pm$4.0\\
V Cen & 5.49392 & 0.289 & 0.000 & 673 & $\pm$20 & 636 & $\pm$48 & 42.0 & $\pm$1.2 & 39.6 & $\pm$2.9\\
CS Vel & 5.904740 & 0.847 & -0.005 & 2908 & $\pm$86 & 3204 & $\pm$297 & 38.5 & $\pm$1.1 & 42.3 & $\pm$4.0\\
BB Sgr & 6.63699 & 0.284 & -0.035 & 789 & $\pm$10 & 729 & $\pm$43 & 49.9 & $\pm$0.6 & 45.6 & $\pm$2.6\\
U Sgr & 6.745226 & 0.403 & 0.000 & 573 & $\pm$6 & 592 & $\pm$21 & 47.7 & $\pm$0.5 & 49.1 & $\pm$1.7\\ \hline
$\eta$ Aql & 7.176779 & 0.149 & 0.000 & 248 & $\pm$5 & 253 & $\pm$15 & 48.4 & $\pm$1.0 & 49.2 & $\pm$2.8\\
S Sge & 8.382086 & 0.127 & -0.010 & 695 & $\pm$11 & 713 & $\pm$25 & 59.1 & $\pm$1.0 & 60.7 & $\pm$2.2\\
S Nor & 9.754244 & 0.189 & 0.000 & 949 & $\pm$14 & 996 & $\pm$51 & 70.7 & $\pm$1.0 & 73.8 & $\pm$3.7\\
Z  Lac & 10.885642 & 0.404 & -0.005 & 1905 & $\pm$38 & 1979 & $\pm$55 & 71.4 & $\pm$1.4 & 73.7 & $\pm$2.1\\
XX Cen & 10.95337 & 0.260 & -0.040 & 1647 & $\pm$17 & 1601 & $\pm$90 & 69.4 & $\pm$0.7 & 67.2 & $\pm$3.7\\
V340 Nor & 11.287 & 0.315 & 0.000 & 1667 & $\pm$142 & 1972 & $\pm$420 & 67.1 & $\pm$5.7 & 79.2 & $\pm$16.9\\
UU Mus & 11.63641 & 0.413 & -0.005 & 3224 & $\pm$125 & 3227 & $\pm$433 & 74.1 & $\pm$2.9 & 73.5 & $\pm$9.9\\
U Nor & 12.643710 & 0.892 & 0.000 & 1325 & $\pm$37 & 1310 & $\pm$68 & 76.3 & $\pm$2.1 & 75.0 & $\pm$3.9\\
BN Pup & 13.67310 & 0.438 & 0.000 & 3802 & $\pm$88 & 4184 & $\pm$277 & 83.3 & $\pm$1.9 & 90.9 & $\pm$6.0\\
LS Pup & 14.146400 & 0.478 & 0.000 & 5010 & $\pm$129 & 5032 & $\pm$438 & 90.2 & $\pm$2.3 & 90.3 & $\pm$7.8\\ \hline
VW Cen & 15.03618 & 0.448 & 0.000 & 3550 & $\pm$64 & 3758 & $\pm$186 & 86.7 & $\pm$1.6 & 91.2 & $\pm$4.5\\
X Cyg & 16.386332 & 0.288 & 0.000 & 1194 & $\pm$10 & 1233 & $\pm$33 & 105.5 & $\pm$0.9 & 108.6 & $\pm$2.8\\
Y Oph & 17.12413 & 0.655 & -0.035 & 573 & $\pm$8 & 590 & $\pm$42 & 89.5 & $\pm$1.2 & 92.1 & $\pm$6.6\\
VY Car & 18.9154892 & 0.243 & -0.020 & 1968 & $\pm$20 & 1958 & $\pm$98 & 112.7 & $\pm$1.1 & 112.2 & $\pm$5.6\\
RY Sco & 20.320144 & 0.777 & 0.000 & 1215 & $\pm$19 & 1253 & $\pm$62 & 99.9 & $\pm$1.6 & 103.1 & $\pm$5.1\\
RZ Vel & 20.3969 & 0.335 & 0.000 & 1571 & $\pm$21 & 1527 & $\pm$64 & 114.8 & $\pm$1.5 & 110.6 & $\pm$4.6\\
WZ Sgr & 21.8496 & 0.467 & 0.000 & 1764 & $\pm$38 & 1800 & $\pm$139 & 121.8 & $\pm$2.6 & 123.3 & $\pm$9.6\\
WZ Car & 23.01320 & 0.384 & 0.000 & 3757 & $\pm$114 & 3678 & $\pm$177 & 112.1 & $\pm$3.4 & 108.5 & $\pm$5.2\\
VZ Pup & 23.17100 & 0.471 & 0.000 & 4027 & $\pm$104 & 3962 & $\pm$243 & 96.9 & $\pm$2.5 & 95.5 & $\pm$5.9\\
SW Vel & 23.443130 & 0.349 & -0.020 & 2459 & $\pm$28 & 2557 & $\pm$85 & 117.3 & $\pm$1.4 & 122.2 & $\pm$4.1\\ \hline
T Mon & 27.03428 & 0.209 & 0.000 & 1439 & $\pm$36 & 1490 & $\pm$55 & 149.6 & $\pm$3.8 & 154.6 & $\pm$5.7\\
RY Vel & 28.129259 & 0.562 & -0.005 & 2458 & $\pm$36 & 2410 & $\pm$185 & 139.9 & $\pm$2.1 & 136.4 & $\pm$10.5\\
AQ Pup & 30.1040 & 0.512 & -0.055 & 3106 & $\pm$64 & 3251 & $\pm$216 & 147.8 & $\pm$3.1 & 154.8 & $\pm$10.4\\
KN Cen & 34.029641 & 0.926 & 0.005 & 4011 & $\pm$83 & 3990 & $\pm$231 & 185.8 & $\pm$3.9 & 184.3 & $\pm$10.8\\
$\ell$ Car & 35.54804 & 0.170 & -0.025 & 561 & $\pm$6 & 559 & $\pm$19 & 179.0 & $\pm$1.8 & 179.9 & $\pm$6.4\\
U Car & 38.81234 & 0.283 & -0.035 & 1497 & $\pm$17 & 1545 & $\pm$68 & 157.6 & $\pm$1.9 & 161.8 & $\pm$7.2\\
RS Pup & 41.4400 & 0.446 & 0.000 & 2004 & $\pm$59 & 1984 & $\pm$122 & 208.1 & $\pm$6.1 & 205.1 & $\pm$12.8\\
SV Vul & 44.994772 & 0.570 & -0.045 & 2553 & $\pm$43 & 2438 & $\pm$149 & 224.0 & $\pm$3.8 & 213.7 & $\pm$13.8
\enddata
\end{deluxetable}

\end{document}